# Rule-based Generation of Diff Evolution Mappings between Ontology Versions

Michael Hartung, Anika Groß, Erhard Rahm

*Department of Computer Science, University of Leipzig, Germany*

{hartung,gross,rahm}@informatik.uni-leipzig.de

***Abstract*— Ontologies such as taxonomies, product catalogs or web directories are heavily used and hence evolve frequently to meet new requirements or to better reflect the current instance data of a domain. To effectively manage the evolution of ontologies it is essential to identify the difference (Diff) between two ontology versions. We propose a novel approach to determine an expressive and invertible diff evolution mapping between given versions of an ontology. Our approach utilizes the result of a match operation to determine an evolution mapping consisting of a set of basic change operations (insert/update/delete). To semantically enrich the evolution mapping we adopt a rule-based approach to transform the basic change operations into a smaller set of more complex change operations, such as merge, split, or changes of entire subgraphs. The proposed algorithm is customizable in different ways to meet the requirements of diverse ontologies and application scenarios. We evaluate the proposed approach by determining and analyzing evolution mappings for real-world life science ontologies and web directories.**

## 1. INTRODUCTION

Ontologies are increasingly used, e.g., to semantically and consistently annotate and categorize objects. In life sciences, large biomedical ontologies such as the Gene Ontology (GO) [5] are used to describe the functions of objects such as genes or proteins, and the ontological information is utilized in many analytical studies. Similarly, product catalogs of online shops, comparison portals or web directories are heavily used to semantically categorize products or websites. Such ontologies are helpful to human users but also for applications, e.g., for focused search queries or website recommendations to relevant subcategories.

Most ontologies evolve frequently to meet new requirements or to better reflect the current instance data of a domain. For instance, product catalogs will be adapted continuously to promote certain kinds of products or to eliminate obsolete categories. As analyzed in [7], biomedical ontologies experience especially frequent changes to better reflect new research insights, e.g., derived from molecular-biological experiments. The newest versions of such ontologies are periodically released.

In this paper we study the problem of determining the so-called diff evolution mapping between two given versions of an ontology. The goal is to find out the changes that have transformed the old to the new version of the ontology. This evolution mapping is informative and significant to users and applications of ontologies to find out how a previously used ontology version has changed. In particular, the evolution mapping can be used to find out whether existing applications would have to be adapted to use the new ontology version. Furthermore, the evolution mapping is needed for migrating instance data.

While there has been a huge amount of research on matching ontologies [14,16], determining the Diff of two ontologies has received only little attention (see section on Related Work). Fortunately, implementing Diff can be based on Match to find corresponding elements of two ontologies that have been changed. Diff also has to consider added and deleted ontology elements which are not represented in match results. While basic diff mappings based on add/delete and change of individual ontology elements (concepts, relationships) are relatively easy to find we observe that such mappings reflect a low-level evolution view that is of limited usefulness for human users especially for large ontologies. We therefore aim at determining semantically more expressive Diff evolution mappings capturing complex ontology changes such as merging, splitting and moving of ontology concepts or adding and deleting entire subgraphs. Such expressive evolution mappings are expected to be useful also for the developers of ontologies who might have performed many small changes that can be represented by a potentially much smaller number of complex changes.

For illustration we use the running example of Figure 1 on the evolution of a fictitious product catalog for storage drives. The goal is to derive the evolution mapping between the two ontology (catalog) versions. A basic Diff approach only supporting add/delete/change operations for individual ontology elements would derive a deletion of the 'DVD-ROM' and 'CD-RW' categories from the old version although these concepts (categories) are merged into concept 'Other'. We propose a more expressive Match-based Diff generation supporting complex changes such as merging concepts. In Figure 1, the names of the white concepts remain unchanged during the evolution and we assume that correspondences between these concepts are part of the match result (though not shown explicitly for clarity of presentation). Dashed lines indicate further matching concepts that are relevant for the Diff mapping. The correspondences help our Diff approach to correctly determine that concepts 'DVD-ROM', 'CD-RW' as well as 'Other' have been merged into 'Other' of the changed version. Further complex changes include the addition of a whole sub-ontology rooted at 'Solid State Disks' and the move of concepts '1.8' and '2½' from 'Hard Disc Drives' to the newly added inner concept 'Notebook'.

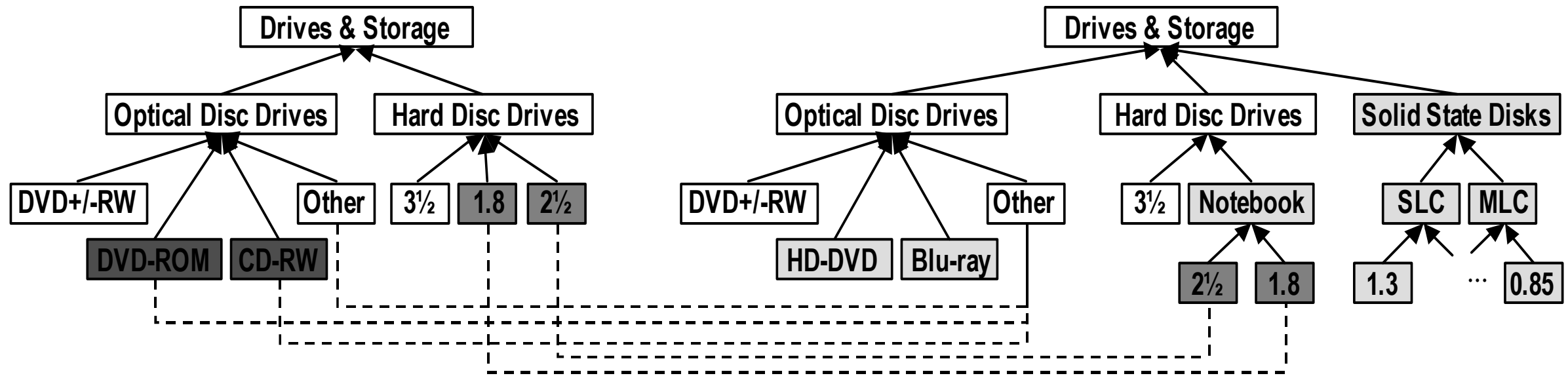

Figure 1. Motivating Example – Evolution of a product catalog for storage drives (left: old version, right: new version)

We propose a rule-based approach for automatically determining such complex changes within a semantically rich Diff evolution mapping. Specifically, we make the following contributions:

- We introduce a model for invertible Diff evolution mappings between ontology versions that is based on a set of basic and complex change operations.
- We present a generic, rule-based Diff algorithm to determine the evolution mapping between given ontology versions based on the result of a match operation (consisting of simple 1:1 correspondences). The evolution mapping is automatically determined by applying a set of so-called COG (change operation generating) rules. We show the correctness of the algorithm and determine its complexity. We also provide an algorithm that uses the evolution mapping to derive the newer ontology version from the older one.
- Our approach also determines the inverse evolution mapping consisting of the inverse of all change operations. This inverse mapping can be used to derive the older ontology version from the newer version.
- We implemented the algorithm and could successfully apply it for different kinds of ontologies. We report on two use cases for determining the Diff between different versions of the Gene Ontology and a web directory.

In this initial study we focus on determining evolution mappings solely at the ontology level and leave the evolution (migration) of ontology instances for future work.

In Section 2 we introduce our ontology model, the considered set of basic and complex changes as well as our model of match mappings and evolution mappings. Section 3 introduces the set of COG rules to determine the change operations that occurred during the evolution. Section 4 outlines the overall approach for determining the Diff evolution mapping based on the introduced COG rules. Furthermore, it shows the correctness of the algorithm and describe the use of the Diff mapping and its inverse for ontology migration. We present evaluation results in Section 5 and discuss related work in Section 6. We finally summarize and outline possibilities for future work.

## 2. MODELS AND PROBLEM STATEMENT

In the following we introduce the models used by our diff algorithm. Particularly, we outline the ontology model as well as change operations, match and evolution mappings. Finally, we state the problem we address.

### 2.1 Ontology Model and Versions

An ontology $O = (C, A, R)$ consists of concepts $C$ having associated attributes of $A$. The concepts are interrelated by directed relationships of $R$ and form a directed acyclic graph (DAG). A special concept called *root* has no relationships to any parent. Each ontology concept has a special attribute called identifier (*id*) which is used to unambiguously identify a concept within the ontology. This identifier can be given, e.g., accession numbers of concepts used in life science ontologies. If not present they need to be generated. For instance, in our running example we can use the unambiguous category names as identifiers. Concepts can have optional concept attributes $a = (a_{concept}, a_{name}, a_{value}) \in A$ which describe the concept in more detail. For instance, life science ontologies use an 'obsolete' attribute for outdated concepts or a 'definition' for description. Note that we do not consider instance-specific attributes which describe the content of instances of an ontology[1].

$R$ consists of a set of directed relationships $r = (r_{source}, r_{type}, r_{target})$ of type $r_{type}$ interconnecting concepts $r_{source}$ and $r_{target}$. The most common relationship type in ontologies is 'is_a' describing an inheritance between two concepts. Other important relationship types are 'part_of' and 'has_parts' denoting the inclusion of a concept in another or vice versa.

An ontology version $O_v = (C_v, A_v, R_v)$ of version $v$ represents a snapshot of an ontology at a specific point in time. The elements of $O_v$ are assumed to be valid until a new ontology version is released. Ontology providers distribute new releases at regular time intervals or whenever a significant number of changes has been incorporated. For instance, the Gene Ontology Consortium daily releases a new version of the popular Gene Ontology.

[1] Our model is motivated by life science ontologies that typically have no directly associated instance data. However, the model is also applicable to other kinds of ontologies.

## 2.2 Change Operations

We consider two sets of change operations for ontology evolution: basic changes (set $B_{op}$) and complex changes (set $C_{op}$). The presented operations are supported by our current design and implementation but can be extended to deal with specific requirements.

*Basic changes* are applied on a single concept, attribute or relationship and deal with either a map (change), addition or deletion resulting in these nine operations in $B_{op}$:

- *mapC*(*c1,c2*): maps a concept *c1* of the first ontology version to a concept *c2* of the second version
- *addC*(*c*): insertion of a new concept *c* in the changed ontology
- *delC*(*c*): deletion of an existing concept *c* from the old ontology version
- *mapR*(*r1,r2*): maps a relationship *r1* of the first ontology version to a relationship *r2* of the second version
- *addR*(*r*): insertion of a new relationship *r*
- *delR*(*r*): deletion of an existing relationship *r*
- *mapA*(*a1,a2*): maps an attribute *a1* of the first to an attribute *a2* of the second ontology version
- *addA*(*a*): addition of an attribute *a*
- *delA(a)*: deletion of an existing attribute *a*

We will later use these basic change operations for ontology version migrations. For instance, we add a new concept with *addC* and specify its position in the ontology structure using an *addR* change.

All other changes are called *complex changes* and belong to $C_{op}$. As we will see they are based on basic changes or other complex changes and thus specify changes at a higher level of abstraction. On the other hand the effect of complex changes can ultimately be expressed by basic changes and we will utilize this for executing an evolution mapping on an ontology version to derive the changed ontology version.

Some of the complex changes operate on single elements (denoted with lower case letters), in particular:

- *substitute*(*c1,c2*): concept *c1* is replaced by *c2*
- *move*(*c, c_from, c_to*): moves a concept *c* and its subgraph from *c_from* under *c_to*
- *toObsolete*(*c*): concept *c* is set obsolete, i.e., it should not be used anymore
- *revokeObsolete*(*c*): the obsolete status of *c* is revoked, i.e., the concept becomes active again

The last two changes are specific to life science ontologies to set and reset an obsolete status of a concept. This illustrates that our evolution approach can cope with such domain-specific change operations and is thus customizable for ontologies from different domains.

Most of our complex changes refer to multiple ontology elements (sets denoted with upper case letters), in particular:

- *addLeaf*(*c, C_Parents*): insertion of a leaf concept *c* below the concepts in *C_Parents*
- *delLeaf*(*c, C_Parents*): deletion of a leaf concept *c* situated below the concepts in *C_Parents*
- *merge*(*Source_C, target_c*): merges multiple source concepts *Source_C* into one target concept *target_c*
- *split*(*source_c, Target_C*): splits one source concept *source_c* into multiple target concepts *Target_C*
- *addSubGraph*(*c_root, C_Sub*): inserts a new subgraph with root *c_root* and concepts *C_Sub*
- *delSubGraph*(*c_root, C_Sub*): removes an existing subgraph with root *c_root* and concepts *C_Sub*

For example, the merge of source categories 'DVD-ROM', 'CD-RW' and 'Other' into target category 'Other' in our running example can be described as

*merge*({DVD-ROM, CD-RW, Other}, Other).

Complex change operations can be implemented by a series of simpler changes and ultimately by basic change operations as we will see. For each complex change we will maintain the simpler changes underlying it according to our rule-based diff generation. We thus maintain lineage information for complex changes as well as dependencies between change operations. In particular, we can thus determine by which basic operations a complex change can be implemented. For instance, our complex merge operation can be derived from three basic *mapC* operations: *mapC*(DVD-ROM, Other), *mapC*(CD-RW, Other) and *mapC*(Other, Other).

Each change operation has an inverse that undoes the effect of the change. The inverse of a change operation *chgOp* typically uses a permutation of *chgOp*'s parameters. For instance, the inverse of *merge(Source_C,target_c)* is *split*(*target_c, Source_C*), i.e., a single source concept is split into multiple target concepts. This symmetry of change operations allows us to derive for every change operation the associated inverse change operation. Appendix A lists all change operations and their inverses.

## 2.3 Match and Evolution Mappings

We represent changes between two ontology versions $O_{old}$ and $O_{new}$ as a mapping. In model management [2,3] a mapping *map*($O_{old}$, $O_{new}$) connects elements of an old ontology version $O_{old}$ with elements of a new version $O_{new}$. We distinguish between a match mapping *match*($O_{old}$, $O_{new}$) and an evolution mapping *diff*($O_{old}$, $O_{new}$). Match mappings represent semantic correspondences between two ontology versions and thus interrelate unchanged elements as well as changed but corresponding (semantically equivalent or related) ontology elements. For our purpose, we only require simple match mappings consisting of 1:1 correspondences between concepts, i.e., *match* ($O_{old}$, $O_{new}$) = {*matchC*(*c1, c2*) | $c1 \in O_{old}$, $c2 \in O_{new}$}.

By contrast an evolution mapping highlights the differences and covers all changes that occurred between two ontology versions, including additions and deletions. Un-changed ontology elements included in a match mapping are not part of an evolution mapping. *Diff mappings* can contain all change operations for single or multi-valued ontology elements *e1*, … as introduced in the previous section: $diff(O_{old}, O_{new}) = \{chgOp(e1,\ldots) \mid chgOp \in B_{Op} \cup C_{Op}\}$. The diff should be *complete*, i.e., it should contain all changes between $O_{old}$ and $O_{new}$. Hence it should also be possible to derive the changed ontology $O_{new}$ by applying the changes of mapping $diff(O_{old}, O_{new})$ to the old version $O_{old}$.

The simplest kind of diff mapping, $diff_{basic}$, only contains basic change operations, i.e., map, add and delete operations: $diff_{basic}(O_{old}, O_{new}) = \{chgOp(e1,\ldots) \mid chgOp \in B_{Op}\}$. However, the main goal is to derive a semantically expressive diff specifying the occurred evolution by complex changes as much as possible. This final $diff_{compact}$ therefore should contain only the semantically most expressive change operations which are not part of any other change operation. As the name suggests, $diff_{compact}(O_{old}, O_{new})$ will generally have fewer operations than the corresponding $diff_{basic}$ since a complex change typically replaces several basic changes.

For evolution mappings $diff(O_{old}, O_{new})$ we also want to determine the inverse evolution mapping that can be used to migrate $O_{new}$ to $O_{old}$. We will use the inverse of the change operations in $diff(O_{old}, O_{new})$ to create the inverse mapping and show that it is equivalent to $diff(O_{new}, O_{old})$.

## 2.4 Problem Statement

The problem that we investigate in this paper is the following. For two given ontology versions $O_{old}$ and $O_{new}$ of the same ontology and a match mapping $match(O_{old}, O_{new})$ the task is to compute the basic evolution mapping $diff_{basic}(O_{old},O_{new})$ and a semantically expressive evolution mapping $diff_{compact}(O_{old},O_{new})$ as well as their inverse mappings. The Diff algorithm should be able to recognize any defined change operation. The diff mapping (and its inverse) should be complete, i.e., contain all changes between the two input versions. Thus, the new (old) ontology version can be constructed from the old (new) ontology version and the (inverse) diff evolution mapping. The algorithms should also be scalable to large ontologies.

# 3. CHANGE OPERATION GENERATING RULES

The identification of basic and complex change operations is based on **C**hange **O**peration **G**enerating **R**ules (COG rules). Each rule is defined by a set of pre-conditions expressed as first-order predicates. If all pre-conditions are fulfilled, a sequence of resulting actions (or a single action) is applied to **create** new change operations or **eliminate** existing ones that are replaced by newly created operations and thus no longer needed. Depending on the type of generated change operation we distinguish between (1) *Basic*, and (2) *Complex* COG rules in the following. We further use (3) *Aggregation* rules to iteratively determine more complex change operations for sets of ontology elements. Rules not only generate changes but may also eliminate redundant change operations from the evolution mapping. The various rules will be used in the Diff generation approach presented in Section 4. Our approach is extensible in a modular way by adding new rules, e.g., to consider specific ontology characteristics or to find further kinds of change operations.

In the following we describe the different types of rules in more detail and provide examples for illustration. Appendix A lists the complete set of our current COG rules. In the rule definitions we denote single elements of an ontology with lower case letters ($a,b,\ldots \in O$), and element sets with upper case letters ($A,B,\ldots \subseteq O$). Each rule has a unique number also indicating the type of rule ($b_1$, $b_2$, … for basic, $c_1$, $c_2$, … for complex, and $a_1$, $a_2$, … for aggregation COG rules). Since rules evaluate existing change operations to derive more expressive changes, there are dependencies between rules leading to a partial order in which rules can be applied (see Algorithm in Section 4). We have chosen the rule numbers to already reflect their execution order, e.g., rule ($b_{10}$) should be applied before ($b_{11}$).

## 3.1 Basic COG Rules

The basic COG rules (b-COG) primarily use information from the match mapping and the ontology versions to determine basic change operations. The following five b-COG rules are used to determine *addC*, *delC* and *mapC* change operations:

($b_1$) $\exists c \in O_{new} \wedge \neg\exists matchC(a,c) \rightarrow \mathbf{create}[addC(c)]$

($b_2$) $\exists c \in O_{old} \wedge \neg\exists matchC(c,a) \rightarrow \mathbf{create}[delC(c)]$

($b_3$) $\exists matchC(a,b) \wedge a \neq b \rightarrow \mathbf{create}[mapC(a,b)]$

($b_4$) $\exists matchC(a,a) \wedge \exists matchC(a,b) \wedge a \neq b \rightarrow \mathbf{create}[mapC(a,a)]$

($b_5$) $\exists matchC(a,a) \wedge \exists matchC(b,a) \wedge a \neq b \rightarrow \mathbf{create}[mapC(a,a)]$

In our running example ($b_1$) determines concept additions (*addC*) such as for 'Blu-ray' and 'HD-DVD'. ($b_3$) creates *mapC* changes that map between changed concepts, e.g., *mapC*(CD-RW, Other). Furthermore, ($b_4$) and ($b_5$) look for concepts that have multiple matches to others and create corresponding *mapC* changes. The existence of multiple correspondences implies a changed semantics for the concept that is expressed in the evolution mapping. Particularly, *mapC*(Other, Other) is created since multiple concepts, e.g., 'CD-RW' and 'DVD-ROM' match with 'Other'.

Appendix A lists six further b-COG rules to determine relationship- and attribute-level changes. In Section 4.2 we use these rules to generate a basic diff evolution mapping $diff_{basic}$.

## 3.2 Complex COG Rules

Complex COG rules are non-recursive and determine the complex changes based on either basic change operations or other complex changes. Complex changes on sets of elements are generally derived in two steps as we will see in Section 4. We first apply c-COG rules to create complex changes on single ontology elements and then use an additional aggregation step (using aggregation rules) to combine several element changes into complex changes on set-valued parameters.

For example, for the complex merge operation we first derive partial merge operations on a single input element using the following c-COG rule:

$(c_7)\ \exists mapC(a,c) \wedge \exists mapC(b,c) \wedge \neg\exists mapC(a,d) \wedge \neg\exists mapC(b,e) \wedge a \neq b \wedge c \neq d \wedge c \neq e \rightarrow$ **create**[*merge*({a},c), *merge*({b},c)], **eliminate**[*mapC*(a,c), *mapC*(b,c)]

The left side of the rule ensures that there exist at least two different source concepts *a* and *b* ($a \neq b$) mapping to the same target concept *c*, and that *a* and *b* have no further maps to other target concepts. The constraints ($c \neq d$, $c \neq e$) describe that the target concept *c* is different from *d* and *e* which are the targets of the two change operations that must not exist. If these pre-conditions are fulfilled we create two element-level merge change operations one from concept *a* into concept *c* and one from *b* into *c*, the corresponding basic changes *mapC*(a,c) and *mapC*(b,c) are eliminated. The element-level merges are further aggregated to complex merge operations by aggregation rules (see below).

For our example *mapC*(DVD-ROM, Other), *mapC*(CD-RW, Other) and *mapC*(Other, Other) would produce change operations *merge*({DVD-ROM}, Other), *merge*({CD-RW}, Other) and *merge*({Other}, Other).

## 3.3 Aggregation Rules

Aggregation COG rules (a-COG) are used to determine all affected elements in set-valued complex change operations. Particularly, several related element-level (or multi-valued) change operations can be aggregated into a combined change operation for a more compact representation. Furthermore, redundant element-level change operations can be eliminated since they are now covered in an aggregated change operation. a-COG rules are recursive to incrementally aggregate elements for a particular change operation. At the same time the redundant/intermediate changes are removed.

For instance, the a-COG rule for *merge* looks as follows:

$(a_3)\ \exists merge(A,c) \wedge \exists merge(B,c) \wedge A \neq B \rightarrow$ **create**[*merge*(A∪B,c)], **eliminate**[*merge*(A,c), *merge*(B,c)]

The rule specifies that two existing merge operations on concept sets *A* and *B* for the same target concept *c* can be combined into a merge on the union $A \cup B$ into *c*. Since the merge from *A* into *c* and from *B* into *c* are now covered by *merge*(A∪B,c) we can eliminate the two previous ones. By iteratively applying the rule (see Section 4) we can increasingly aggregate the input sets of the operations until no further aggregation is possible. In our example we would first create *merge*({DVD-ROM, CD-RW}, Other) and eliminate *merge*({DVD-ROM}, Other) and *merge*({CD-RW}, Other). Afterwards the final merge operation *merge*({DVD-ROM, CD-RW, Other}, Other) is created and *merge*({Other}, Other) as well as *merge*({DVD-ROM, CD-RW }, Other) are removed.

# 4. DIFF COMPUTATION

In this section we present our approach to generate a diff evolution mapping. We first give an overview of the approach and discuss how to obtain the match mapping needed as input. In Section 4.3 we describe the algorithm to determine the basic evolution mapping $diff_{basic}$ as well as the final evolution mapping $diff_{compact}$. Furthermore, we prove the correctness of the algorithm and describe the use of the Diff mapping and its inverse for ontology migration.

## 4.1 Overview

The main phases of our Diff approach are illustrated in Figure 2. The input are two versions of the same ontology ($O_{old}$, $O_{new}$). Optionally, background knowledge (e.g., dictionaries, change descriptions) can be provided for matching the ontologies. The result is an expressive diff evolution mapping $diff_{compact}(O_{old}, O_{new})$ and a corresponding $diff_{basic}(O_{old}, O_{new})$ containing only basic change operations. The basic diff mapping can be used for migrating the ontologies. Optionally we determine the inverse evolution mapping from the Diff result (Section 4.6). The complete algorithm operates on a working repository which stores the ontology versions as well as intermediate and final mappings.

The first phase is a *Matching* of the two ontology versions to identify common as well as modified but corresponding ontology elements. The result is a match mapping $match(O_{old}, O_{new})$ consisting of a set of *matchC* correspon-dences.

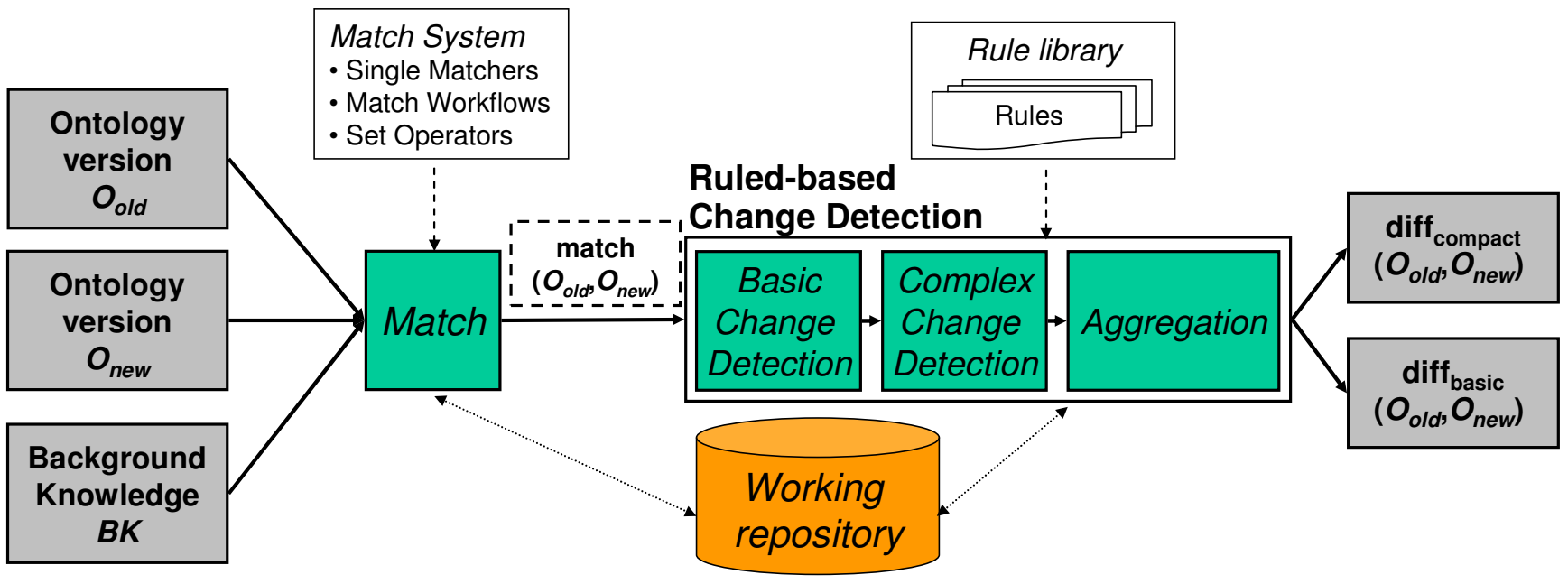

Figure 2. Schematic overview of the rule-based diff approach

The following steps are based on the match result and are completely automatic. They use the COG rules to determine the Diff evolution mapping according to Algorithm 1. It sequentially applies the different kinds of COG rules for finding basic changes (*Basic Change Detection*), complex changes (*Complex Change Detection*), and aggregated complex changes (*Aggregation*).

## 4.2 Matching Phase

The matching phase uses both input versions as well as optional background knowledge to compute a match mapping. Matching ontology versions is typically much easier than matching independent ontologies. This is due to the fact that a new version evolves from the older version and hence a larger part of the old version usually remains unchanged. For life science ontologies, we can further utilize the fact that ontology concepts typically have unambiguous identifiers (accession numbers) so that correspondences are easily determined. For other kinds of ontologies, we can use existing match systems such as COMA++ [1] supporting a variety of match techniques for improved match quality. Furthermore, matching is only semi-automatic, i.e., a user will verify proposed match correspondences and correct them if necessary. Our Diff algorithm will thus assume that the match step provides the correct and complete set of correspondences.

The result of the matching phase for our motivating example in Figure 1 is the following. All white categories exhibiting the same label in the old and new version do match as well as category pairs connected by a dashed line.

## 4.3 Rule-based Change Detection

Algorithm 1 implements the rule-based generation of diff evolution mappings.

**Algorithm 1** *diffEvolMapGen* $(O_1, O_2, M, R)$

**Input:** two ontology versions $O_1$ and $O_2$, match mapping
$M = match(O_1,O_2)$, list of rules $R = [R_{b\text{-}COG},R_{c\text{-}COG},R_{a\text{-}COG}]$
**Output:** diff evolution mappings $diff_{basic}(O_1,O_2)$, $diff_{compact}(O_1,O_2)$

1: $diff_{basic}(O_1,O_2) \leftarrow diffBasicGen(O_1, O_2, M, R_{b\text{-}COG})$
2: $D \leftarrow diff_{basic}(O_1,O_2)$
3: **for each** $r \in R_{c\text{-}COG}$ **do**
4: $\quad D \leftarrow applyRule\ (D, r)$
5: **end for**
6: $diff_{compact}(O_1,O_2) \leftarrow applyAggRules(D, R_{a\text{-}COG})$
7: **return** $[diff_{basic}(O_1,O_2), diff_{compact}(O_1,O_2)]$

Its input are two ontology versions $O_1/O_2$, a match mapping $M$ between $O_1$ and $O_2$ as well as the list of COG rules $R$. The result contains two diff evolution mappings namely a basic diff evolution mapping $diff_{basic}(O_1,O_2)$ and a semantically enriched one, $diff_{compact}(O_1,O_2)$. In the three main steps of the algorithm the three kinds of COG rules are applied to generate the respective changes.

Step 1 calls a procedure *diffBasicGen* to generate the basic diff evolution mapping $diff_{basic}(O_1, O_2)$ based on match mapping $M$ and the list of all b-COG rules ($R_{b\text{-}COG}$).

**Algorithm 2** *diffBasicGen* $(O_1, O_2, M, R_{b\text{-}COG})$

**Input:** two ontology versions $O_1$ and $O_2$, match mapping
$M = match(O_1,O_2)$, list of b-COG rules $R_{b\text{-}COG}$
**Output:** basic diff evolution mapping $D = diff_{basic}(O_1,O_2)$

1: $D \leftarrow$ empty
2: **for each** $r \in R_{b\text{-}COG}$ **do**
3: $\quad D \leftarrow applyBasicRule\ (D, r, O_1, O_2, M)$
4: **end for**
5: **return** $D$

The b-COG rules need only to be applied once (*applyBasicRule*) in a pre-defined order (see numbering in Appendix A). *applyBasicRule* evaluates all $O_1/O_2$ elements whether they are relevant for the respective rule and creates the corresponding changes. We will later prove the correctness of *diffBasicGen* (see Section 4.4) and use $diff_{basic}$ for ontology migration purposes (see Section 4.5).

The processing of c-COG rules (steps 2-5 of Algorithm 1) is similar to the processing of b-COG rules in that each rule needs to be applied only once in the predetermined order. Rule processing starts with the basic diff evolution mapping and iteratively enriches the mapping with complex changes. Furthermore, the eliminate statements in the c-COG rules lead to the removal of changes that are replaced by newly generated complex changes.

The *Aggregation* step requires to apply the a-COG rules multiple times to recursively aggregate set-valued change operations. This functionality is realized by the *applyAggRules* procedure (Algorithm 3) called in step 6 of Algorithm 1.

**Algorithm 3** *applyAggRules* $(D, R_{a\text{-}COG})$

**Input:** diff evolution mapping $D$ $(D' \leftarrow D)$, list of rules $R_{a\text{-}COG}$
**Output:** diff evolution mapping $D'$

1: **do**
2: $\quad D \leftarrow D'$
3: $\quad$**for each** $r \in R_{a\text{-}COG}$ **do**
4: $\quad\quad D' \leftarrow applyRule(D', r)$
5: $\quad$**end for**
6: **while** $(D \neq D')$
7: **return** $D'$

Algorithm 3 accepts an intermediate diff evolution mapping $D$ and a sorted list of a-COG rules $R_{a\text{-}COG}$ as input. In each iteration (do-while loop) the rules of $R_{a\text{-}COG}$ are applied in their pre-defined order (see numbering in Appendix A). Thus, we can apply a-COG rules multiple times (once per iteration) to recursively detect and aggregate multiple change operations. The application of one rule (*applyRule*) modifies the temporary evolution mapping $D'$ according to the rule's resulting actions (**create**, **eliminate**) if the preconditions are fulfilled and the inferred change operations do not exist yet. We apply rules as long as new changes are inferred and the temporary mapping changes ($D \neq D'$).

Appendix B contains the results of running the algorithm *diffEvolMapGen* for the complete running example. For illustration, we explain in the following the generation of a complex change, namely the addition of the 'Solid State Disks' subgraph in the running example (right-hand side of Figure 1).

For our subgraph example the b-COG addition rule ($b_1$) would detect five concept additions: *addC*(Solid State Disks), *addC*(SLC), *addC*(MLC), *addC*(1.3) and *addC*(0.85). For detecting subgraph additions the following c-COG rules are applied:

($c_5$) $\exists addC(a) \wedge \neg\exists addR(r) \wedge a{=}r_{target} \wedge \exists addR(s) \wedge a{=}s_{source}$
$\rightarrow$ **create**[*addLeaf*(a,{$s_{target}$})], **eliminate**[*addC*(a), *addR*(s)]

($c_9$) $\exists addC(a) \wedge \exists addLeaf(b,B) \wedge a \in B$
$\rightarrow$ **create**[*addSubGraph*(a,{b})],
**eliminate**[*addC*(a), *addLeaf*(b,B)]

The first rule ($c_5$) is used to detect leaf concept additions. Rule ($c_9$) is based on the results of ($c_5$) and infers subgraph additions connecting a newly added concept *a* and a leaf concept *b* rooted at *a*. In our example '1.3' and '0.85' are classified as leaf concept additions. Afterwards rule ($c_9$) infers *addSubGraph*(SLC, {1.3}) and *addSubGraph*(MLC, {0.85}). We then can apply the following a-COG rules:

($a_5$) $\exists addSubGraph(a,A) \wedge \exists addC(b) \wedge \exists addR(r) \wedge r_{source}{=}a \wedge r_{target}{=}b$
$\rightarrow$ **create**[*addSubGraph*(b,{a}$\cup$A)],
**eliminate**[*addSubGraph*(a,A), *addC*(c), *addR*(r)]

($a_6$) $\exists addSubGraph(a,A) \wedge \exists addSubGraph(a,B) \wedge A{!=}B$
$\rightarrow$ **create**[*addSubGraph*(a,A$\cup$B)],
**eliminate**[*addsubGraph*(a,A), *addSubGraph*(a,B)]

Particularly, ($a_5$) recursively aggregates added concepts into larger subgraphs. If multiple subgraph additions with the same root exist, we can aggregate these into one by fusing their sub concepts ($a_6$). In our example ($a_5$) would detect two changes: *addSubGraph*(Solid State Disks,{SLC, 1.3}) and *addSubGraph*(Solid State Disks, {MLC, 0.85}) which are finally aggregated into *addSubGraph*(Solid State Disks, {SLC, 1.3, MLC, 0.85}) by ($a_6$).

The final *diff* result is the semantically richest evolution mapping which cannot further be compacted w.r.t. the used set of rules. For our subgraph example the only remaining change is *addSubGraph*(Solid State Disks, {SLC, 1.3, MLC, 0.85}). All other changes have been eliminated during rule application. Appendix B shows that $diff_{compact}$ for our complete running example merely contains six complex changes while the basic evolution mapping $diff_{basic}$ has 25 changes.

The *complexity* of Algorithm 1 depends on the number of rule applications and the number of changed ontology elements. b-COG and c-COG rules are applied only once for all affected ontology elements. For an ontology size of *n* elements, the execution cost for these rules thus is at most O(*n*) if we assume that the number of changed elements is proportional to *n*. The recursive a-COG rules incrementally reduce at least two change operations into one, resulting in O(log(*n*)) rule activations and a total complexity of O(*n* log(*n*)). Given that typically only a small fraction of the *n* elements is affected by evolutionary changes we expect fast execution times of our algorithm even for large ontologies.

## 4.4 Correctness

We now show that the proposed algorithm for generating Diff evolution mappings is correct, in particular that it generates all changes and that it terminates. We first show that the generation of the basic diff evolution mapping is complete, i.e., determines all basic changes between two input ontology versions $O_{old}$ and $O_{new}$. We focus on concept changes; the correctness proof for relationship and attribute changes is analogous.

**Theorem 1:** *The b-COG rules applied in diffBasicGen generate a complete basic diff evolution mapping* $diff_{basic}(O_{old},O_{new})$ *containing*

a) *all concept additions (addC) between* $O_{old}$ *and* $O_{new}$
b) *all concept deletions (delC) between* $O_{old}$ *and* $O_{new}$
c) *all concept changes (mapC) including concepts that map to multiple concepts in the other ontology version.*

To prove the theorem, we refer to the five b-COG rules ($b_1$-$b_5$) introduced in Section 3.1 and applied in *diffBasicGen*. The rules distinguish between concepts that match with at least one concept in the other ontology version and those that do not match. For all non-matching concepts of $O_{new}$ b-COG rule ($b_1$) generates *addC* change operations. b-COG rule ($b_2$) generates concept deletions (*delC*) for all non-matching concepts of $O_{old}$. Matching concepts occur in correspondences $matchC(a,b) \in match(O_{old},O_{new})$ and are processed by b-COG rules ($b_3$), ($b_4$) and ($b_5$). Rule ($b_3$) creates a *mapC(a,b)* change if *a* and *b* are unequal ($a \neq b$). For ($a{=}b$), rules ($b_4$) and ($b_5$) ensure that we only create a *mapC* change if the concept is involved in further correspondences, i.e., has not remained the same. Hence, all *matchC*(*a*,*a*) connecting unchanged concepts are not included in $diff_{basic}$. In summary, $diff_{basic}$ reflects all basic changes but does not relate unchanged ontology parts.

To prove the completeness of $diff_{compact}$ we have to show that algorithm *diffEvolMapGen* produces the semantically richest evolution mapping for the given c-COG/a-COG rules. The completeness is guaranteed by three facts. First, the input is $diff_{basic}$ which is complete (see Theorem 1). Second, the rules cover all considered changes and are iteratively applied for all possible input configurations as long as new changes can be derived. Third, *diffEvolMapGen* terminates as shown below.

We finally show that the algorithm *diffEvolMapGen* terminates. This is mainly ensured by two facts. First, all rules operate on a finite number of ontology elements in $O_{old}/O_{new}$ and do not create new ontology elements. Second, the evaluation of all rules terminates since we apply them only once or as long as the mapping changes. So, the application of b-COG and c-COG rules terminates since they are non-recursive and are applied only once based on a pre-defined order. a-COG rules are recursive but always reduce the number of change operations by aggregating ontology elements. Particularly, each rule uses ≥2 input change operations which are fused into one; the input change operations are eliminated. This steady reduction of change operations terminates when the most aggregated change operations have been found.

When adding further change operations and corresponding COG rules the correctness of our Diff approach is preserved by observing the characteristics of the existing rule set. In particular, no cyclic dependencies between change rules should be introduced and only aggregation rules reducing the number of changes may be recursive.

## 4.5 Ontology Version Migration

An important application of diff evolution mappings is the migration of ontology versions. We can migrate an old version $O_1$ to the changed version $O_2$ by applying the basic diff evolution mapping $diff_{basic}(O_1,O_2)$ on $O_1$. This approach results in an in-place ontology version that retains the unchanged ontology elements. Changes are thus limited to the ontology parts affected by the evolution supporting an efficient migration.

Algorithm 4 (*ontVersionMig*) implements the migration of ontology version $O_1$ to $O_1'$ based on the basic diff evolution mapping $diff_{basic}(O_1,O_2)$.

**Algorithm 4** *ontVersionMig* $(O_1, D)$

**Input:** ontology version $O_1$, basic diff evolution mapping $D = diff_{basic}(O_1,O_2)$

**Output:** migrated ontology version $O_1'$

```
O1' ← O1
performOrder ← [delA, delR, delC, mapC, mapA, mapR,
                addC, addA, addR]
for each chgOp in performOrder do
    O1' ← perform(D.getChgOp(chgOp), O1')
end for
return O1'
```

It executes the basic change operations of $diff_{basic}$ in a pre-defined *performOrder*. For the deletions, we first remove concept attributes. We then remove the concepts from the ontology structure and finally eliminate themselves. For the *map* changes we first need to substitute concepts (*mapC*), afterwards *mapR* and *mapA* can be executed, e.g., we update a changed attribute value or relationship type. Finally, for *additions* we first add the concept and then its attributes and relationships.

We now show the correctness of the algorithm.

**Theorem 2:** *Algorithm ontVersionMig is correct, i.e., for the basic diff evolution mapping* $diff_{basic}(O_1,O_2)$ *determined by algorithm diffBasicGen, it creates the new ontology version* $O_2$ *from the original version* $O_1$.

We prove the theorem for concept changes. We need to show that the generated ontology version is complete, i.e., it (1) contains all concepts and (2) contains no further/other concepts. First, it is easy to see that the algorithm removes deleted concepts indicated by *delC* changes so that they do not become part of $O_2$. Analogously, the concepts of the domain of *mapC* changes are eliminated and thus do not appear in $O_2$. Second, an unchanged concept *c* already available in $O_1$ should be present in $O_2$ as well. Particularly, such a *c* is not covered by any change operation of $diff_{basic}$ and thus remains in $O_2$. Finally, concepts specified by *addC* changes are added to the unchanged ontology part and thus become part of $O_2$. In the same way the range concepts of *mapC* changes in $diff_{basic}$ are inserted.

## 4.6 Inverse Diff Mappings

Inverse diff evolution mappings can be applied to undo the changes in an evolution, i.e., we want to switch back from a changed ontology version to the old one. Our change model allows an easy way to determine an inverse diff evolution mapping because every change operation has an unique inverse change operation as introduced in Section 2 and listed in Appendix A. Hence, we can simply replace every change operation by its inverse change operation to obtain the inverse mapping.

**Theorem 3:** *The inverse of a basic diff evolution mapping* $diff_{basic}(O_1,O_2)$ *is correct, i.e., is identical to* $diff_{basic}(O_2,O_1)$.

We prove this theorem for concept changes. We will show that the inverse of every change operation in $diff_{basic}(O_1,O_2)$ is in $diff_{basic}(O_2,O_1)$ and also that $diff_{basic}(O_2,O_1)$ does not contain additional changes. The addition of a concept *c* *addC*(c) in $diff_{basic}(O_1,O_2)$ has *delC*(c) as inverse. Since *c* is not in $O_1$ but in $O_2$ b-COG rule ($b_2$) will create a *delC*(c) change in $diff_{basic}(O_2,O_1)$. Analogously, ($b_1$) would create *addC*(c) in $diff_{basic}(O_2,O_1)$ if *c* is in $O_1$ but not in $O_2$ which corresponds to a *delC*(c) change (inverse of *addC*(c)) in $diff_{basic}(O_1,O_2)$. A *mapC*(a,b) change in $diff_{basic}(O_1,O_2)$ has *mapC*(b,a) as its inverse and, according to rules ($b_3$,$b_4$,$b_5$), requires a correspondence *matchC*(a,b). Changing the domain and range leads to a *matchC*(b,a) correspondence and thus to a *mapC*(b,a) change in $diff_{basic}(O_2,O_1)$. The concept changes in $diff_{basic}(O_2,O_1)$ are only created by rules ($b_1$) to ($b_5$) like those of $diff_{basic}(O_1,O_2)$. Hence, there can be no further changes in addition to the changes in the inverse of $diff_{basic}(O_1,O_2)$.

The inverse of $diff_{basic}(O_1, O_2)$ gives us a basic evolution mapping that, using the algorithm *ontVersionMig*, can be used to correctly migrate from $O_2$ to $O_1$.

To evaluate and verify the proposed algorithms we apply a roundtrip migration between two ontology versions as illustrated in Figure 3. For this we first migrate version $O_1$ to a changed version $O_1$' as follows: $O_1' = ontVersionMig(O_1, diff_{basic}(O_1, O_2))$. Due to the completeness of the migration algorithm, $O_1'$ equals $O_2$. Second, we determine the inverse of $diff_{basic}(O_1, O_2)$, i.e., $diff_{basic}(O_2, O_1)$, and use it to migrate $O_1'$: $O_1'' = ontVersionMig(O_1', diff_{basic}(O_2, O_1))$. The resulting ontology $O_1''$ will equal $O_1$ (completeness of migration algorithm and correctness of inverse). The round-trip migration has been performed on several real-world ontologies as discussed in the next section.

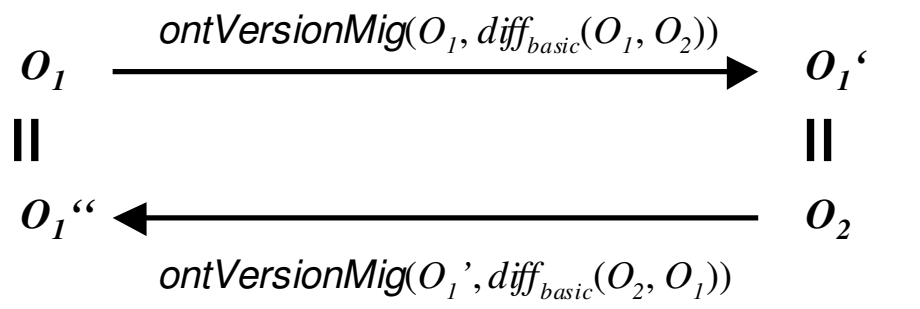

Figure 3. Roundtrip migration scenario

# 5. Evaluation Experiments

We determine and evaluate evolution mappings for two real-world ontologies namely the Gene Ontology (GO) [5] and a portion of the DMoz web directory (www.dmoz.org) related to soccer. GO is widely used in bioinformatics for the uniform annotation of molecular-biological objects such as proteins or genes. The DMoz web directory is an open directory project and represents a comprehensive human directory of the Web. It is also used by popular portals and search engines such as Google or Lycos. We consider two yearly evolution periods (three ontology versions) for both GO and DMoz: 2008 (2008-01→2009-01) and 2009 (2009-01→2010-01). For matching the GO versions we use the unambiguous accession numbers to identify corresponding concepts in a *match* mapping. For DMoz soccer we utilized a semi-automatic matching which incorporates label and root path (context) of a category. GO contains concepts, is-a and part-of relationships and attributes. DMoz is a pure is-a hierarchy and has no concept attributes. Hence, we only consider concepts and relationships in the following analyses. The implementation of the proposed Diff approach is in Java utilizing a MySQL database as working repository.

Table 1 lists the ontology and mapping sizes for the considered evolution scenarios. GO is much larger than DMoz and increased its size substantially in the two years while DMoz shrinked especially in the first year (2008). The last columns show the sizes of *match* and $diff_{compact}$. The results indicate that the algorithm produces a relatively compact diff evolution mapping compared to the size of the new version in all scenarios. For instance $diff_{compact}(GO_{2009\text{-}01}, GO_{2010\text{-}01})$ is about factor 20 smaller than the whole new version $GO_{2010\text{-}01}$. For DMoz the compactness of the diff is even more significant, e.g., factor 93 for $diff_{compact}(DMoz_{2009\text{-}01}, DMoz_{2010\text{-}01})$. These observations indicate that our algorithm finds compact evolution mappings highlighting only the changes between two ontology versions. Hence, ontology providers could release such a compact and expressive but still complete evolution mapping instead of a whole version.

Table 1. Statistics for GO and DMoz Scenarios

| $O_{old}$ - $O_{new}$ | $\vert O_{old}\vert$ | $\vert O_{new}\vert$ | \|match\| | $\vert diff_{compact}\vert$ |
|---|---|---|---|---|
| **$GO_{2008\text{-}01}$ - $GO_{2009\text{-}01}$** | 66,121 | 75,180 | 25,774 | 8,450 |
| **$GO_{2009\text{-}01}$ - $GO_{2010\text{-}01}$** | 75,180 | 84,714 | 27,861 | 4,284 |
| **$DMoz_{2008\text{-}01}$ - $DMoz_{2009\text{-}01}$** | 6,683 | 6,185 | 3,130 | 255 |
| **$DMoz_{2009\text{-}01}$ - $DMoz_{2010\text{-}01}$** | 6,185 | 6,079 | 3,047 | 65 |

Table 2. Distribution of Change Operations

| | GO | | DMoz | |
|---|---|---|---|---|
| | **2008** | **2009** | **2008** | **2009** |
| **add** | 4,187 | 1,355 | 0 | 0 |
| **del** | 1,407 | 316 | 0 | 0 |
| **map** | 0 | 0 | 0 | 0 |
| **addLeaf** | 768 | 796 | 18 | 6 |
| **delLeaf** | 0 | 0 | 199 | 46 |
| **merge** | 70 | 83 | 16 | 4 |
| **move** | 1,499 | 1,200 | 0 | 0 |
| **substitute** | 0 | 1 | 15 | 9 |
| **toObsolete** | 225 | 66 | 0 | 0 |
| **addSubGraph** | 294 | 467 | 0 | 0 |
| **delSubGraph** | 0 | 0 | 7 | 0 |
| $\Sigma$ | **8,450** | **4,284** | **255** | **65** |

The diff computation took about 150 sec for GO and merely 5 sec for DMoz (2009). The rule application required 9 iterations for GO and 3 for DMoz. This is a consequence of the smaller size of DMoz and the rare occurrence of complex change operations such as the addition of large subgraphs.

We investigated the evolution mappings in more detail, we show the absolute frequency of basic and complex change operations in Table 2. We observe significant differences between both ontologies. GO's increasing size is reflected by a high number of information extending operations such as *addLeaf* operations as well as a significant amount of subgraph additions. For instance, the largest subgraph (GO:0070887, 'cellular response to chemical stimulus') added in 2009 encompasses 94 new concepts. Note that no concepts have been deleted in GO since it merely marks concepts as obsolete if they are no longer required or outdated; thus outdated information remains in GO for compatibility reasons. On the other hand, DMoz experienced mainly deletions but only few additions and substitutions indicating that the considered subontology soccer has mainly been consolidated. Interestingly, as for our running example, the final diff mapping of DMoz contains no basic change operations, i.e., all basic operations could be covered by complex changes.

We further compare the sizes of $diff_{basic}$ and the enhanced $diff_{compact}$ which is separated into basic and complex changes. Table 3 also shows the size ratio $|diff_{compact}|/|diff_{basic}|$ which is between 31 and 54%. For DMoz all basic changes could be eliminated from the final *diff*, for GO the number of complex changes was also almost twice as high than the number of remaining basic changes (2009).

Table 3. $diff_{basic}$ vs. $diff_{compact}$

| | GO | | DMoz | |
|---|---|---|---|---|
| | **2008** | **2009** | **2008** | **2009** |
| **$\vert diff_{basic}\vert$** | 15,781 | 13,504 | 630 | 149 |
| **$\vert diff_{compact}\vert$** | 8,450 | 4,284 | 255 | 65 |
| ╚ **#basic** | 5,594 | 1,671 | 0 | 0 |
| ╚ **#complex** | 2,856 | 2,613 | 255 | 65 |
| **ratio in %** | **53.5%** | **31.7%** | **40.5%** | **43.6%** |

TABLE 4. ROUND TRIP MIGRATION RESULTS

| diff($O_1$,$O_2$) | \|$O_1$\| | \|$diff_{basic}$\| | \|$O_1 \cap O_1''$\| | \|$O_1 \cup O_1''$\| |
|---|---|---|---|---|
| $GO_{2008\text{-}01}$ - $GO_{2009\text{-}01}$ | 200,169 | 29,944 | 200,169 | 200,169 |
| $GO_{2009\text{-}01}$ - $GO_{2010\text{-}01}$ | 218,176 | 33,555 | 218,176 | 218,176 |
| $DMoz_{2008\text{-}01}$ - $DMoz_{2009\text{-}01}$ | 6,683 | 630 | 6,683 | 6,683 |
| $DMoz_{2009\text{-}01}$ - $DMoz_{2010\text{-}01}$ | 6,185 | 149 | 6,185 | 6,185 |

Finally, we performed the roundtrip migration experiment (see Section 4.6) for the two ontologies. To compare the ontologies $O_1$ with $O_1''$ we utilize their representation as sets of concepts, relationships and attributes as introduced in Section 2.1. By testing $O_1 \cap O_1'' = O_1 \cup O_1''$ we ensure that the migrated ontology version $O_1''$ contains exactly the same elements (concepts, relationships, attributes) as $O_1$, indicating the completeness of the evolution mapping and its inverse mapping. The results are presented in the Table 4. The first two columns show the number of elements in the source version (|$O_1$|) and the size of the basic evolution mapping |$diff_{basic}(O_1,O_2)$|, including concepts, relationships, and attributes. The last two columns in the table indicate the number of elements in the intersection and union of $O_1$ with $O_1''$. We could verify that $O_1$ and $O_1''$ always contain exactly the same elements (concepts, relationships, attributes) confirming that the algorithms were correctly implemented thus achieving the completeness properties for the evolution mappings and their inverse mappings.

# 6. RELATED WORK

Most related to our work are previous approaches on change detection for ontology evolution (see [4] for a survey). Approaches can be classified into *incremental* and *direct* ones. Incremental approaches [12] are based on a version log of changes. Thus, they are limited to users who have access to a version log of an ontology. Furthermore, the occurrence of redundant changes in a log makes change detection between two versions difficult. Hence, it is more promising to follow a direct comparison (including a match) to get the diff between two ontology versions. In [9] a fixed-point algorithm PromptDiff as part of the Protégé tool [8] is proposed. The algorithm integrates several heuristic matchers to detect the structural diff between two versions presented in a difference table using only simple add/del/map changes. The algorithm terminates if no more changes are found. The authors of [10] describe an approach to detect high-level changes in RDF/S knowledge bases. They use additions/deletions of triples and logical formula to find high-level changes such as moveClass. Furthermore, so-called heuristic changes such as comment or label modifications can be detected in an optional matching step. Change detection has also been studied for other data models, e.g., structured data. The LaDiff algorithm [17] uses change operations insert, update, delete and move to find a minimum cost edit script for transforming versions of ordered trees. An extended algorithm called MH-DIFF presented in [18] works on unordered trees considering more rich operations such as copy.

In contrast to this previous work we adopt a two-phase diff algorithm to find expressive evolution mappings. First matching is used to determine an initial match mapping connecting corresponding concepts in the input ontology versions. Second the phased and iterative application of rules on the match mapping allows to successively determine more and more complex ontology changes, e.g., we can find changes on entire subgraphs. Due to the distinction between matching and rule application we can flexibly adapt our algorithm to work with different ontologies in various application scenarios, e.g., detecting toObsolete changes in life science ontologies.

Schema evolution (see [13] for a bibliography) has been extensively studied in the area of databases, e.g., for relational databases [15] and XML [6]. The focus of most approaches is on data translation in case of a schema change. For instance, Clio [11] uses logical mappings on schema level which are mapped to transformation queries such as SQL (relational case) or XQuery/XSLT (XML case) to translate dependent data. Our approach focuses on determining an evolution mapping for ontologies that are primarily used to describe instances managed separately, e.g., protein data sources use GO to semantically describe the functions of proteins. Thus, we do not study the transformation of dependent data but our evolution mapping may help users of such sources to adapt their data accordingly.

Diff is also one of the proposed operators in model management, in particular to describe evolution scenarios [2], [3]. In this context, Diff is used to determine the changed information in an evolved model M' compared to the old model M. Specifically, for a changed model M' and the match mapping between M and M' the Diff operator determines the (sub-) model M'' containing the new parts of M' as well as a mapping describing the overlap between M' and M''. Our diff definition, by contrast, returns an evolution mapping consisting of all change operations describing the evolution including rich changes such as merging and splitting of concepts. Furthermore, we provide a rule-based diff implementation for ontologies which is also driven by a match mapping.

# 7. SUMMARY

We presented a new rule-based approach to determine an expressive and invertible diff evolution mapping between two ontology versions. The diff evolution mapping covers basic and complex changes. The approach is based on a match mapping between ontology versions and utilizes Change Operation Generating Rules (COG rules) to find the basic as well as complex change operations. The rules also specify which simpler changes are replaced by more expressive changes. The evaluation on real-world ontologies from life sciences and web directories showed that our Diff approach generates semantically expressive and rela-tively compact evolution mappings. We were further able to use the evolution mappings and their inverse mappings to roundtrip migrate between ontology versions. Our approach can be customized for different domains by providing additional change operations and COG rules.

We see several possibilities for future work. First we want to apply our evolution mappings especially the complex change operations in advanced application scenarios. For instance, in ontology matching we may use the evolution mapping to migrate outdated ontology mappings to be consistent with newer ontology versions. This would save the effort to rematch complete ontologies and previous match results could be reused. Furthermore, we will also study the migration of ontology instances based on diff evolution mappings.

## ACKNOWLEDGEMENTS

This work was supported by German Research Foundation (DFG) via grant RA 497/18-1 ('Evolution of Ontologies and Mappings').

## APPENDIX A

### LIST OF CHANGE OPERATIONS, INVERSES AND RULES

The table lists all change operations introduced in Section 2 and their inverses. Furthermore, the table includes a full list of the corresponding COG rules (types b-COG, c-COG, a-COG) described in Section 3 and used in the rule-based change detection phase of our diff algorithm.

| Change operations | Inverse of change operations | COG-Rules | Rule-Type |
|---|---|---|---|
| **Basic change operations** | | | |
| ***addC* (c)** | ***delC* (c)** | $\exists c\in O_{new} \wedge\neg\exists \mathit{matchC}(a,c)\rightarrow\mathbf{create}[\mathit{addC}(c)]$ | $\mathbf{b_1}$ |
| ***delC* (c)** | ***addC* (c)** | $\exists c\in O_{old} \wedge\neg\exists \mathit{matchC}(c,a)\rightarrow\mathbf{create}[\mathit{delC}(c)]$ | $\mathbf{b_2}$ |
| ***mapC* (c2,c1)** | ***mapC* (c1,c2)** | $\exists \mathit{matchC}(a,b)\wedge a\neq b\rightarrow\mathbf{create}[\mathit{mapC}(a,b)]$ | $\mathbf{b_3}$ |
| | | $\exists \mathit{matchC}(a,a)\wedge\exists \mathit{matchC}(a,b)\wedge a\neq b\rightarrow\mathbf{create}[\mathit{mapC}(a,a)]$ | $\mathbf{b_4}$ |
| | | $\exists \mathit{matchC}(a,a)\wedge\exists \mathit{matchC}(b,a)\wedge a\neq b\rightarrow\mathbf{create}[\mathit{mapC}(a,a)]$ | $\mathbf{b_5}$ |
| ***addR* (r)** | ***delR* (r)** | $\exists r\in O_{new} \wedge\neg\exists r\in O_{old}\rightarrow\mathbf{create}[\mathit{addR}(r)]$ | $\mathbf{b_6}$ |
| ***delR* (r)** | ***addR* (r)** | $\exists r\in O_{old} \wedge\neg\exists r\in O_{new}\rightarrow\mathbf{create}[\mathit{delR}(r)]$ | $\mathbf{b_7}$ |
| ***mapR* (r2,r1)** | ***mapR* (r1,r2)** | $\exists \mathit{delR}(r)\wedge\exists \mathit{addR}(s)\wedge r_{source}=s_{source}\wedge r_{target}=s_{target}\wedge r_{type}\neq s_{type}$ $\rightarrow\mathbf{create}[\mathit{mapR}(r,s)]$, $\mathbf{eliminate}[\mathit{delR}(r), \mathit{addR}(s)]$ | $\mathbf{b_8}$ |
| ***addA* (a)** | ***delA* (a)** | $\exists p\in O_{new} \wedge\neg\exists p\in O_{old}\rightarrow\mathbf{create}[\mathit{addA}(p)]$ | $\mathbf{b_9}$ |
| ***delA* (a)** | ***addA* (a)** | $\exists p\in O_{old} \wedge\neg\exists p\in O_{new}\rightarrow\mathbf{create}[\mathit{delA}(p)]$ | $\mathbf{b_{10}}$ |
| ***mapA* (a2,a1)** | ***mapA* (a1,a2)** | $\exists \mathit{delA}(p)\wedge\exists \mathit{addA}(q)\wedge p_{concept}=q_{concept}\wedge p_{name}=q_{name}\wedge p_{value}\neq q_{value}$ $\rightarrow\mathbf{create}[\mathit{mapA}(p,q)]$, $\mathbf{eliminate}[\mathit{delA}(p), \mathit{addA}(q)]$ | $\mathbf{b_{11}}$ |
| **Complex change operations** | | | |
| ***substitute* (c1,c2)** | ***substitute* (c2,c1)** | $\exists \mathit{mapC}(a,b)\wedge\neg\exists \mathit{mapC}(a,c)\wedge\neg\exists \mathit{mapC}(d,b)\wedge a\neq b\wedge a\neq d\wedge b\neq c$ $\rightarrow\mathbf{create}[\mathit{substitute}(a,b)]$, $\mathbf{eliminate}[\mathit{mapC}(a,b)]$ | $\mathbf{c_1}$ |
| ***move* (c,c_to,c_from)** | ***move* (c,c_from,c_to,)** | $\exists \mathit{delR}(r)\wedge\exists \mathit{addR}(s)\wedge r_{source}=s_{source}\wedge r_{target}\neq s_{target}\wedge r_{type}=s_{type}$ $\rightarrow\mathbf{create}[\mathit{move}(r_{source},r_{target},s_{target})]$, $\mathbf{eliminate}[\mathit{delR}(r), \mathit{addR}(s)]$ | $\mathbf{c_2}$ |
| ***toObsolete* (c)** | ***revokeObsolete* (c)** | $\exists \mathit{mapA}(p,q)\wedge p_{concept}=q_{concept}\wedge p_{name}=\text{'obsolete'}\wedge p_{name}=q_{name}\wedge p_{value}=\text{'false'}$ $\wedge q_{value}=\text{'true'}\rightarrow\mathbf{create}[\mathit{toObsolete}(p_{concept})]$, $\mathbf{eliminate}[\mathit{mapA}(p,q)]$ | $\mathbf{c_3}$ |
| ***revokeObsolete* (c)** | ***toObsolete* (c)** | $\exists \mathit{mapA}(p,q)\wedge p_{concept}=q_{concept}\wedge p_{name}=\text{'obsolete'}\wedge p_{name}=q_{name}$ $\wedge p_{value}=\text{'true'}\wedge q_{value}=\text{'false'}\rightarrow\mathbf{create}[\mathit{revokeObsolete}(p_{concept})]$, $\mathbf{eliminate}[\mathit{mapA}(p,q)]$ | $\mathbf{c_4}$ |
| ***addLeaf* (c,C_Parents)** | ***delLeaf* (c,C_Parents)** | $\exists \mathit{addC}(a)\wedge\neg\exists \mathit{addR}(r)\wedge a=r_{target}\wedge\exists \mathit{addR}(s)\wedge a=s_{source}$ $\rightarrow\mathbf{create}[\mathit{addLeaf}(a,\{s_{target}\})]$, $\mathbf{eliminate}[\mathit{addC}(a), \mathit{addR}(s)]$ | $\mathbf{c_5}$ |
| | | $\exists \mathit{addLeaf}(a,A)\wedge\exists \mathit{addLeaf}(a,B)\wedge A\neq B\rightarrow\mathbf{create}[\mathit{addLeaf}(a,A\cup B)]$, $\mathbf{eliminate}[\mathit{addLeaf}(a,A), \mathit{addLeaf}(a,B)]$ | $\mathbf{a_1}$ |
| ***delLeaf* (c,C_Parents)** | ***addLeaf* (c,C_Parents)** | ∃*delC* (a)∧¬∃*delR* (r)∧a=$r_{target}$∧∃*delR* (s)∧a=$s_{source}$ →**create**[*delLeaf* (a,{$s_{target}$})], **eliminate**[*del* C(a), *delR* (s)] | **$c_6$** |
| | | ∃*delLeaf* (a,A)∧∃*delLeaf* (a,B)∧A≠B→**create**[*delLeaf* (a,A∪B)], **eliminate**[*delLeaf* (a,A), *delLeaf* (a,B)] | **$a_2$** |
| ***merge* (Target_c,source_c)** | ***split* (source_c,Target_c)** | ∃*mapC* (a,c)∧∃*mapC* (b,c)∧¬∃*mapC* (a,d)∧¬∃*mapC* (b,e)∧a≠b∧c≠d∧c≠e →**create**[*merge* ({a},c), *merge* ({b},c)], **eliminate**[*mapC* (a,c), *mapC* (b,c)] | **$c_7$** |
| | | ∃*merge* (A,c)∧∃*merge* (B,c)∧A≠B→**create**[*merge* (A∪B,c)], **eliminate**[*merge* (A,c), *merge* (B,c)] | **$a_3$** |
| ***split* (source_c,Target_c)** | ***merge* (Target_c,source_c)** | ∃*mapC* (c,a)∧∃*mapC* (c,b)∧¬∃*mapC* (d,a)∧¬∃*mapC* (e,b)∧a≠b∧c≠d∧c≠e →**create**[*split* (c,{a}), *split* (c,{b})], **eliminate**[*mapC* (c,a), *mapC* (c,b)] | **$c_8$** |
| | | ∃*split* (c,A)∧∃*split* (c,B)∧A≠B→**create**[*split* (c,A∪B)], **eliminate**[*split* (c,A), *split* (c,B)] | **$a_4$** |
| ***addSubGraph* (c_root,C_sub)** | ***delSubGraph* (c_root,C_sub)** | ∃*addC* (a)∧∃*addLeaf* (b,B)∧a∈B→**create**[*addSubGraph* (a,{b})], **eliminate**[*addC* (a), *addLeaf* (b,B)] | **$c_9$** |
| | | ∃*addSubGraph* (a,A)∧∃*addC* (b)∧∃*addR* (r)∧$r_{source}$=a∧$r_{target}$=b →**create**[*addSubGraph* (b,{a}∪A)],**eliminate**[*addSubGraph* (a,A), *addC* (b), *addR* (r)] | **$a_5$** |
| | | ∃*addSubGraph* (a,A)∧∃*addSubGraph* (a,B)∧A≠B→**create**[*addSubGraph* (a, A∪B)], **eliminate**[*addSubGraph* (a,A), *addSubGraph* (a,B)] | **$a_6$** |
| | | ∃*addSubGraph* (a,A)∧∃*addSubGraph* (b,B)∧∃*addR* (r)∧$r_{source}$=a∧($r_{target}$=b ∨ $r_{target}$ ∈ B) →**create**[*addSubGraph* (b,{a}∪A∪B)], **eliminate**[*addSubGraph* (a,A), *addSubGraph* (b,B), *addR* (r)] | **$a_7$** |
| ***delSubGraph* (c_root,C_sub)** | ***addSubGraph* (c_root,C_sub)** | ∃*delC* (a)∧∃*delLeaf* (b,B)∧a∈B→**create**[*delSubGraph* (a,{b})], **eliminate**[*delC* (a), *delLeaf* (b,B)] | **$c_{10}$** |
| | | ∃*delSubGraph* (a,A)∧∃*delC* (b)∧∃*delR* (r)∧$r_{source}$=a∧$r_{target}$=b →**create**[*delSubGraph* (b,{a}∪A)], **eliminate**[*delSubGraph* (a,A), *delC* (b), *delR* (r)] | **$a_8$** |
| | | ∃*delSubGraph* (a,A)∧∃*delSubGraph* (a,B)∧A≠B→**create**[*delSubGraph* (a, A∪B)], **eliminate**[*delSubGraph* (a,A), *delSubGraph* (a,B)] | **$a_9$** |
| | | ∃*delSubGraph* (a,A)∧∃*delSubGraph* (b,B)∧∃*delR* (r)∧$r_{source}$=a∧($r_{target}$=b ∨ $r_{target}$ ∈ B) →**create**[*delSubGraph* (b,{a}∪A∪B)], **eliminate**[*delSubGraph* (a,A), *delSubGraph* (b,B), *delR* (r)] | **$a_{10}$** |

# APPENDIX B

## RESULTS OF DIFF ALGORITHM FOR RUNNING EXAMPLE

This section presents a run of our diff algorithm on the running example illustrated in Figure 1. The match mapping *match* contains the following correspondences: *matchC*(Drives & Storage, Drives & Storage), *matchC*(Optical Disc Drives, Optical Disc Drives), *matchC*(DVD+/-RW, DVD+/-RW), *matchC*(Other, Other), *matchC*(3½, 3½), *matchC*(2½, 2½), *matchC*(1.8, 1.8), *matchC*(DVD-ROM, Other), *matchC*(CD-RW, Other). The tables below show which COG-rules are applied to generate the diff evolution mappings $diff_{basic}$ and $diff_{compact}$. Greyly marked operations were first created and later eliminated, i.e., only the white operations are in $diff_{compact}$.

| Created by rule | Generated change operation | Eliminated by rule |
|---|---|---|
| **Application of basic COG-Rules** | | |
| **$b_1$** | *addC* (HD-DVD) | **$c_5$** |
| **$b_1$** | *addC* (Blu-ray) | **$c_5$** |
| **$b_1$** | *addC* (Notebook) | |
| **$b_1$** | *addC* (Solid State Disks) | **$a_5$** |
| **$b_1$** | *addC* (SLC) | **$c_9$** |
| **$b_1$** | *addC* (MLC) | **$c_9$** |
| **$b_1$** | *addC* (1.3) | **$c_5$** |
| **$b_1$** | *addC* (0.85) | **$c_5$** |
| **$b_3$** | *mapC* (DVD-ROM, Other) | **$c_7$** |
| **$b_3$** | *mapC* (CD-RW, Other) | **$c_7$** |
| **$b_5$** | *mapC* (Other, Other) | **$c_7$** |
| **$b_6$** | *addR* (HD-DVD, subCatOf, Optical Disc Drives) | **$c_5$** |
| **$b_6$** | *addR* (Blu-ray, subCatOf, Optical Disc Drives) | **$c_5$** |
| **$b_6$** | *addR* (Notebook, subCatOf, Hard Disc Drives) | |
| **$b_6$** | *addR* (1.8, subCatOf, Notebook) | **$c_2$** |
| **$b_6$** | *addR* (2½, subCatOf, Notebook) | **$c_2$** |
| **$b_6$** | *addR* (Solid State Disks, subCatOf, Drives & Storage) | |
| **$b_6$** | *addR* (SLC, subCatOf, Solid State Disks) | **$a_5$** |
| **$b_6$** | *addR* (MLC, subCatOf, Solid State Disks) | **$a_5$** |
| **$b_6$** | *addR* (1.3, subCatOf, SLC) | **$c_5$** |
| **$b_6$** | *addR* (0.85, subCatOf, MLC) | **$c_5$** |
| **$b_7$** | *delR* (1.8, subCatOf, Hard Disc Drives) | **$c_2$** |
| **$b_7$** | *delR* (2½, subCatOf, Hard Disc Drives) | **$c_2$** |
| **$b_7$** | *delR* (DVD-ROM, subCatOf, Optical Disc Drives) | |
| **$b_7$** | *delR* (CD-RW, subCatOf, Optical Disc Drives) | |
| **Application of complex COG-Rules** | | |
| **$c_2$** | *move* (1.8, Hard Disc Drives, Notebook) | |
| **$c_2$** | *move* (2½, Hard Disc Drives, Notebook) | |
| **$c_5$** | *addLeaf* (HD-DVD, {Optical Disc Drives}) | |
| **$c_5$** | *addLeaf* (Blu-ray, {Optical Disc Drives}) | |
| **$c_5$** | *addLeaf* (1.3, {SLC}) | **$c_9$** |
| **$c_5$** | *addLeaf* (0.85, {MLC}) | **$c_9$** |
| **$c_7$** | *merge* ({DVD-ROM},Other) | **$a_3$** |
| **$c_7$** | *merge* ({CD-RW},Other) | **$a_3$** |
| **$c_7$** | *merge* ({Other},Other) | **$a_3$** |
| **$c_9$** | *addSubGraph* (SLC, {1.3}) | **$a_5$** |
| **$c_9$** | *addSubGraph* (MLC, {0.85}) | **$a_5$** |
| **Application of aggregate COG-Rules** | | |
| **$a_3$** | *merge* ({DVD-ROM, CD-RW},Other) | **$a_3$** |
| **$a_3$** | *merge* ({DVD-ROM, CD-RW, Other}, Other) | |
| **$a_5$** | *addSubGraph* (Solid State Disks, {SLC,1.3}) | **$a_6$** |
| **$a_5$** | *addSubGraph* (Solid State Disks, {MLC,0.85}) | **$a_6$** |
| **$a_6$** | *addSubGraph* (Solid State Disks, {SLC,1.3,MLC,0.85}) | |